\documentclass[twocolumn,showpacs,preprintnumbers,amsmath,amssymb,prb]{revtex4-1}
\usepackage[active]{srcltx}
\usepackage{amsmath,amssymb}
\usepackage{epsfig}
\usepackage{graphicx}
\usepackage{bm}
\usepackage{xcolor}

\newcommand{\e}{{\bm e}}

\newcommand{\B}{{\bm B}}

\renewcommand{\j}{{\bm j}}
\renewcommand{\k}{{\bm{k}}}

\def\gsim{\lower.35em\hbox{$\stackrel{\textstyle>}{\textstyle\sim}$}}

\begin{document}
\title{Change of chirality at magic angles of twisted bilayer graphene}

\author{T. Stauber$^{1,2}$, J. Gonz\'alez$^{3}$, and G. G\'omez-Santos$^{4}$}

\affiliation{$^{1}$Materials Science Factory,
Instituto de Ciencia de Materiales de Madrid (ICMM-CSIC), E-28049 Madrid, Spain\\
$^{2}$Institute for Theoretical Physics, University of Regensburg, D-93040 Regensburg, Germany\\
$^{3}$Instituto de Estructura de la Materia, CSIC, E-28006 Madrid, Spain\\
$^{4}$Departamento de F\'{\i}sica de la Materia Condensada, Instituto Nicol\'as Cabrera and Condensed
Matter Physics Center (IFIMAC), Universidad Aut\'onoma de Madrid, E-28049 Madrid, Spain}
\date{\today}

\begin{abstract}
We derive a simple formula for the real-space chirality of twisted bilayer graphene that can be related to the cross-product of its sheet currents. This quantity shows well-defined plateaus for the first remote band as function of the gate voltage which are approximately quantized for commensurate twist angles. The zeroth plateau corresponds to the first magic angle where a sign change occurs due to an emergent $C_6$-symmetry. Our observation offers a new definition of the magic angle based on a macroscopic observable which is accessible in typical transport experiments. 
\end{abstract}

\maketitle
{\it Introduction.} Magic-angle graphene, i.e., twisted bilayer graphene\cite{Lopes07,Shallcross08,Suarez10,Schmidt10,Li10,Trambly10,Bistritzer11,Dean13} (TBG) at an angle of $\theta\sim1.08^\circ$, has attracted tremendous attention since the discovery of correlated insulator states\cite{Kim17,Cao18a} as well as superconductivity.\cite{Cao18b,Yankowitz19,Lu19} Moreover, it represents the first material where Coulomb interactions can drive a system into a topologically non-trivial state leading to a valley and spin-polarised Chern band displaying anomalous ferromagnetism.\cite{Sharpe19,Bultinck19,ZhangMao19} This makes TBG and also related systems such as ABC-trilayer gaphene on a misaligned BN-substrate\cite{Chen19,Serlin20} an ideal platform to study the interplay between correlations and topology.  
\begin{figure}[h]
\includegraphics[width=0.98\columnwidth]{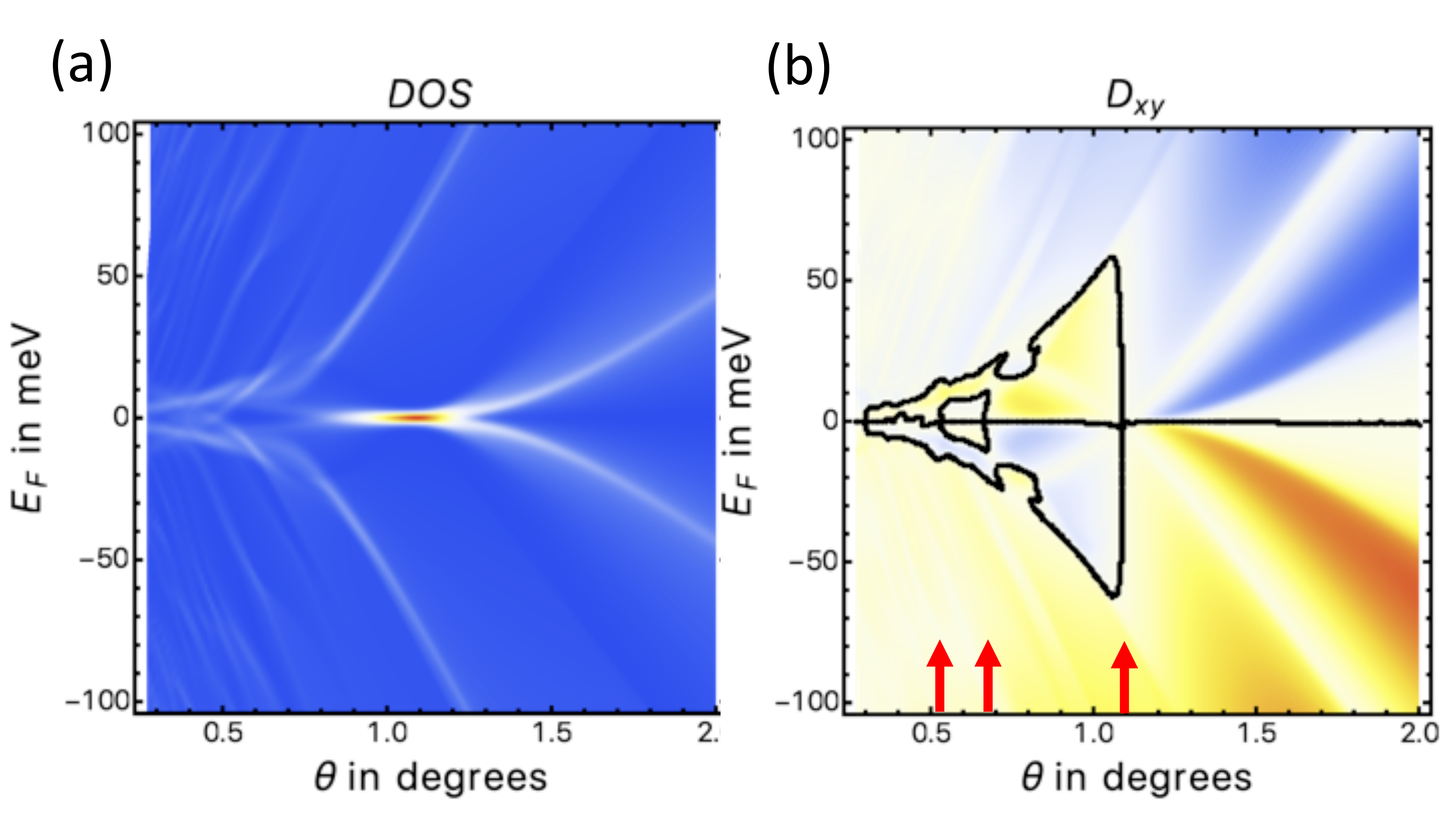}
\includegraphics[width=0.98\columnwidth]{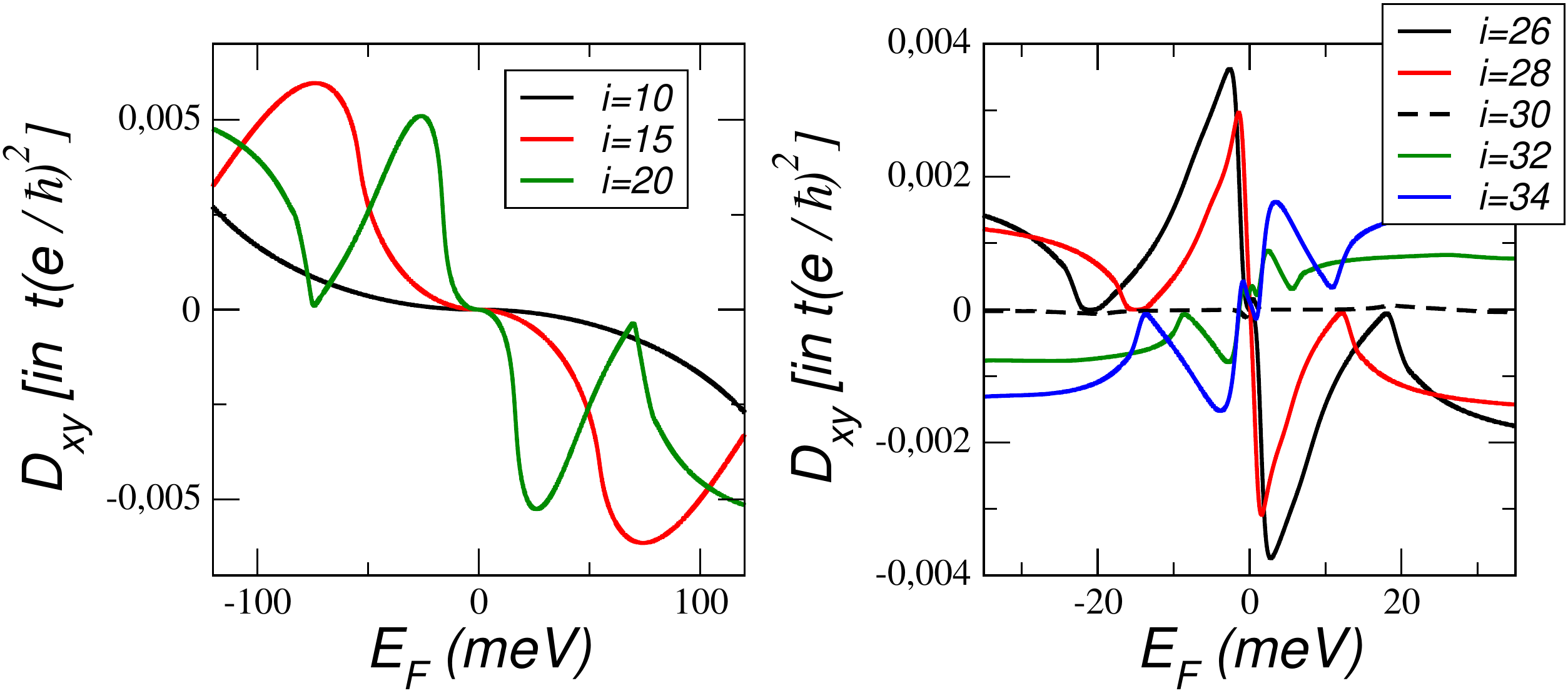}
\caption{\label{Fig1} {\it Upper panels:} Density of States (left) and chiral Drude weight (right) as function of the twist angle $\theta$ and Fermi energy $E_F$. For $D_{xy}$, there are sign changes indicated by the black line. In addition to the horizontal line at $E_F=0$, approximate vertical lines define magic angles where the chirality sign changes for finite range of $E_F$. These angles are signaled by arrows in the corresponding axis. {\it Lower panels:} The chiral Drude weight in units of $t(e/\hbar)^2$ for large twist angles with $i=10,15,20$ (left) and for small twist angles with $i=26,28,30,32,34$ (right) as function of the Fermi energy $E_F$. Twist angles are parametrised by $\cos\theta_i=\frac{3i^2+3i+1/2}{3i^2+3i+1}$ and $t=2.78$eV denotes the in-plane hopping parameter.}
\end{figure}

The focus of this Rapid Communication is on TBG's intrinsic chirality that gives rise to circular dichroism in the absence of symmetry-breaking fields.\cite{Kim16,Suarez17,Addison19,Ochoa20} Opposite enantiomers also show different transport behavior, i.e., the adiabatic application of an in-plane magnetic field provokes a longitudinal current whose direction depends on the sign of the chiral Drude weight $D_{xy}$.\cite{Stauber18,Stauber18b,Bahamon20} For the continuum model,\cite{Lopes07,Bistritzer11} this response is related to the cross-product of the sheet current densities
\begin{align}
\label{HallDrude}
D_{xy}=\frac{1}{2A}\sum_{\k,n}\e_z\cdot(\j_{\k,n}^1\times\j_{\k,n}^2)\delta(\epsilon_{\k,n}-E_F)\;,
\end{align}
where $\j_{\k,n}^\ell=\langle \k,n|\j^\ell|\k,n\rangle$ and $\epsilon_{\k,n}$ and $|\k,n\rangle$ denote the eigenvalues and eigenvectors, respectively, with $\k$ inside the first Brillouin zone. $A$ labels the area of the sample and $\j_i^\ell$ is the current operator of layer $\ell=1,2$. 

Eq. (\ref{HallDrude}) is the main result of this Rapid Communication and will be derived below. It allows for an easy evaluation and further provides an intuitive interpretation of electronic chirality. A slightly modified formula which evolves the cross-product between one sheet current and the total current would also hold for a general tight-binding model.\cite{TB} 

Due to an approximate particle-hole symmetry, $D_{xy}$ changes sign at the neutrality point, but interestingly, there are also sign-changes at finite carrier density that extend over a large energy range for fixed twist angle, see Fig. \ref{Fig1}. $D_{xy}$ thus defines a set of magic angles that is not limited to the flat band, but also extends to the remote bands. 

\begin{figure*}
\includegraphics[width=0.434\columnwidth]{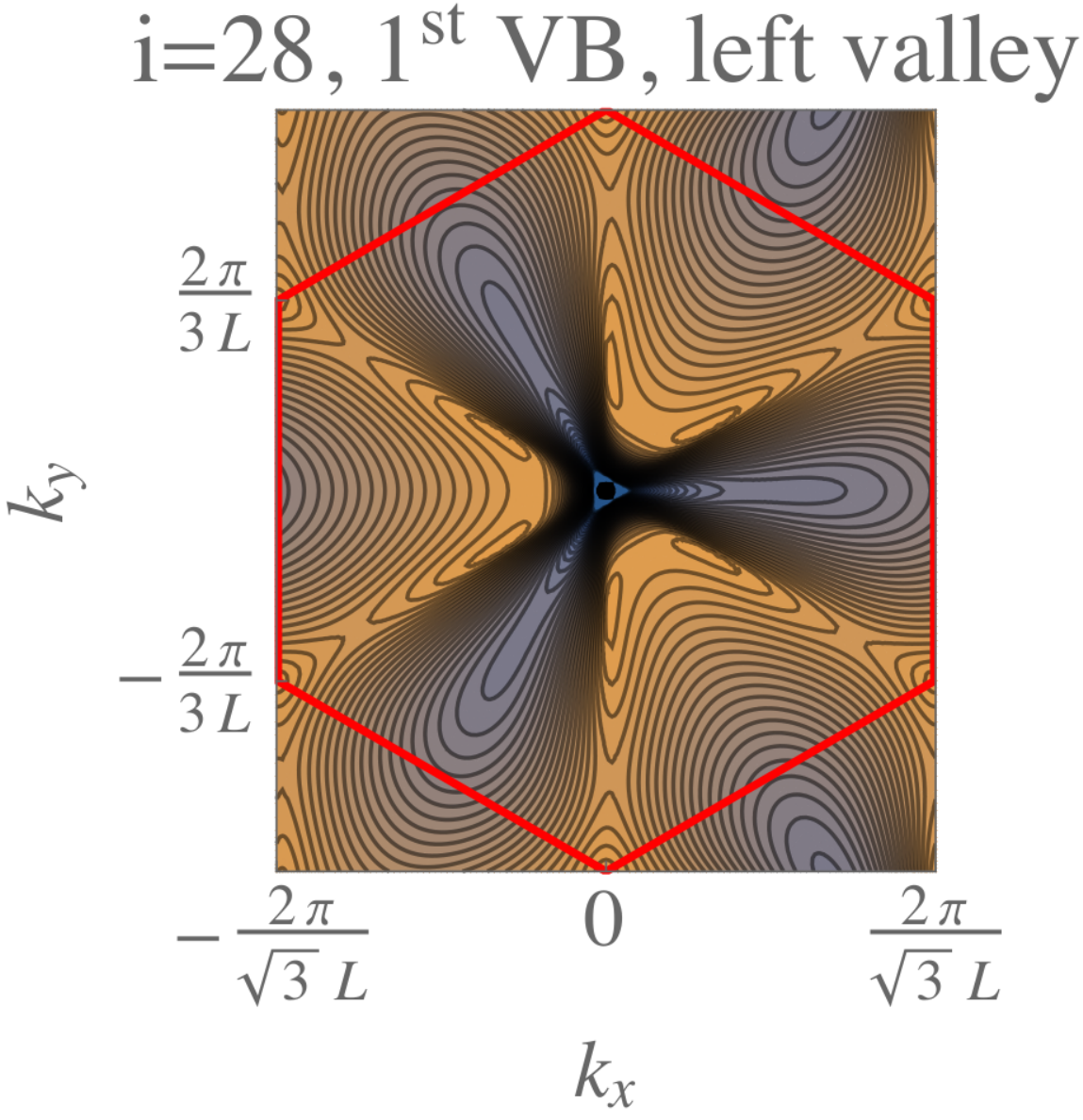}
\includegraphics[width=0.438\columnwidth]{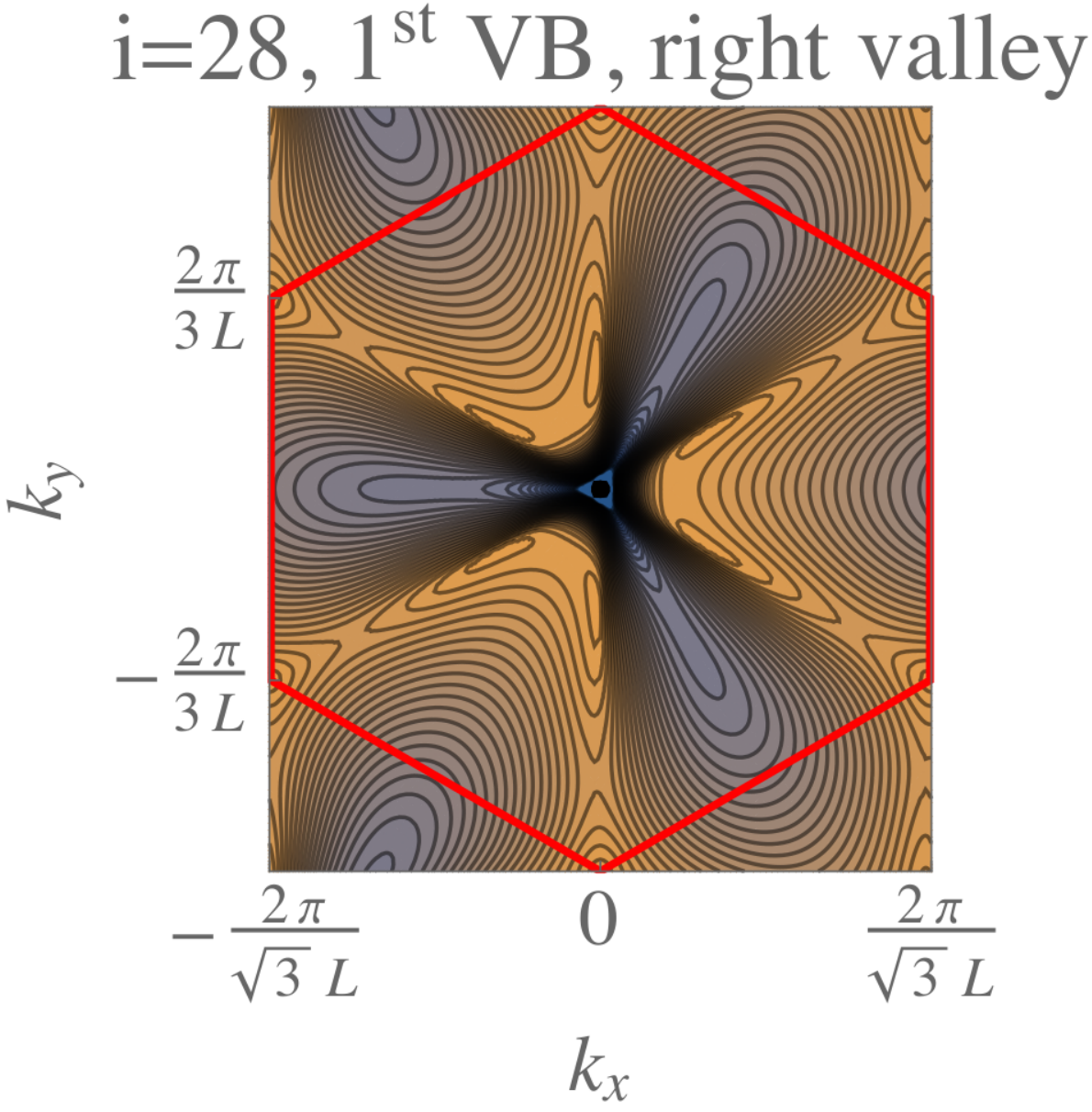}
\includegraphics[width=0.434\columnwidth]{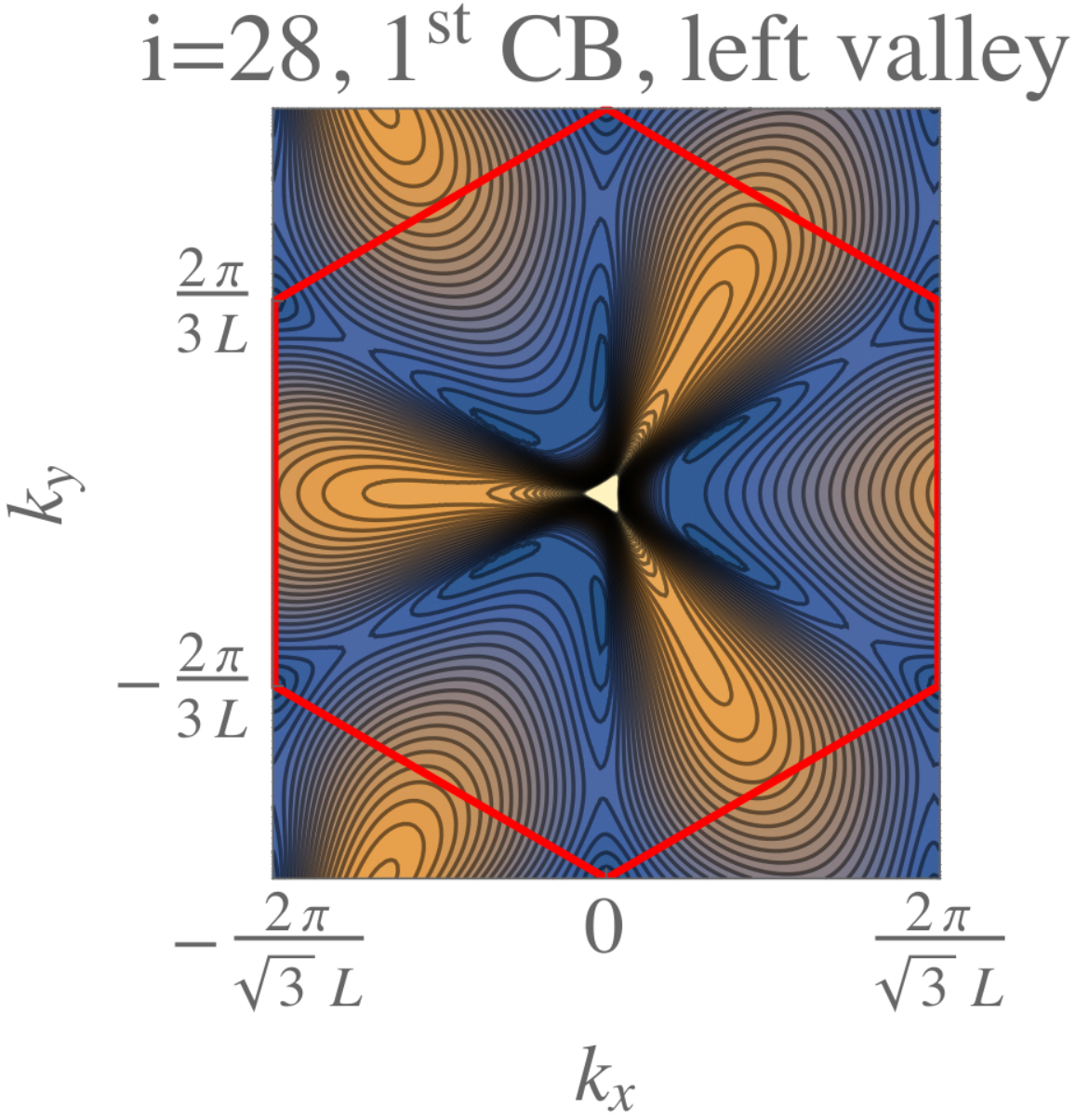}
\includegraphics[width=0.438\columnwidth]{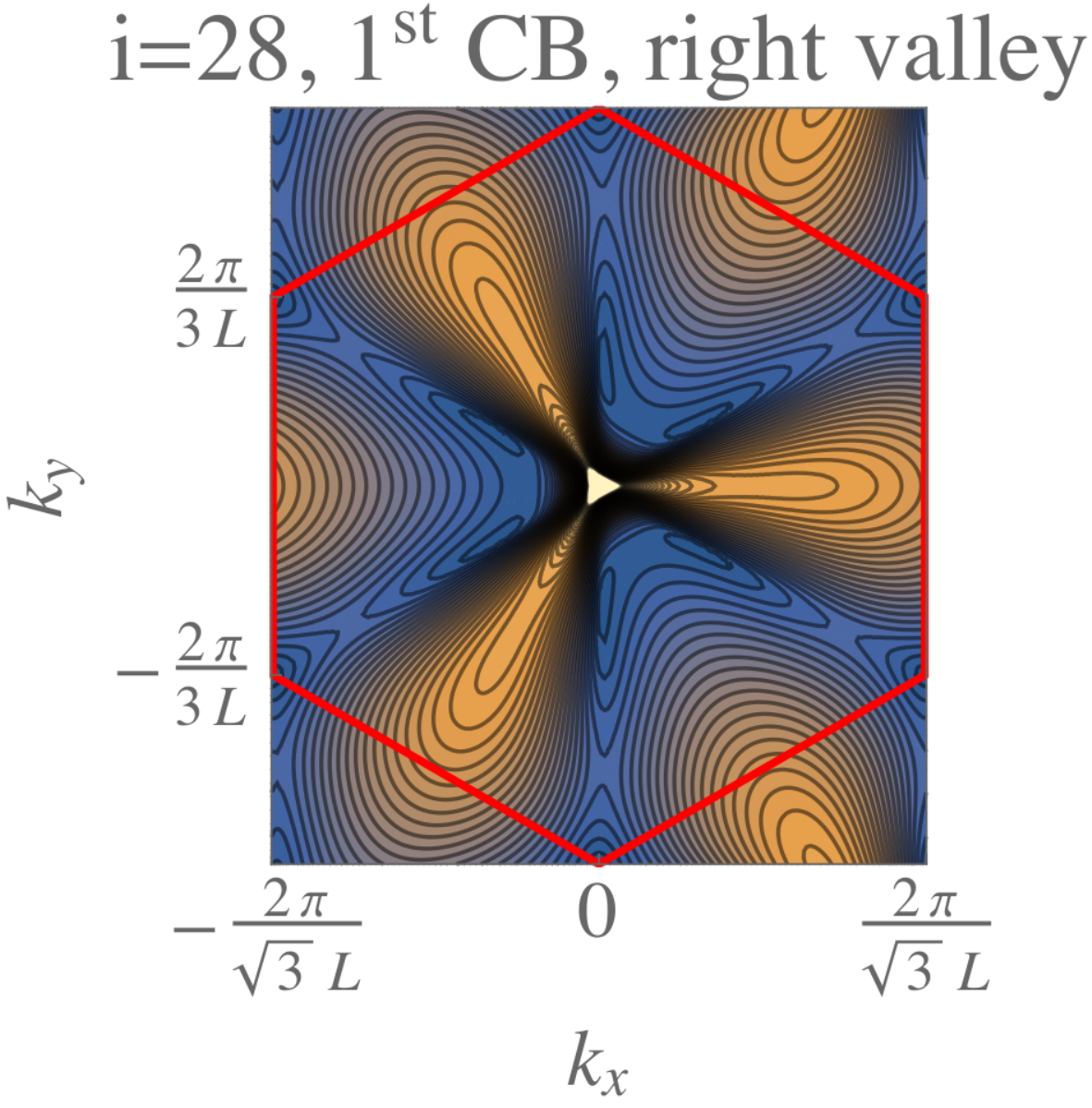}
\includegraphics[width=0.435\columnwidth]{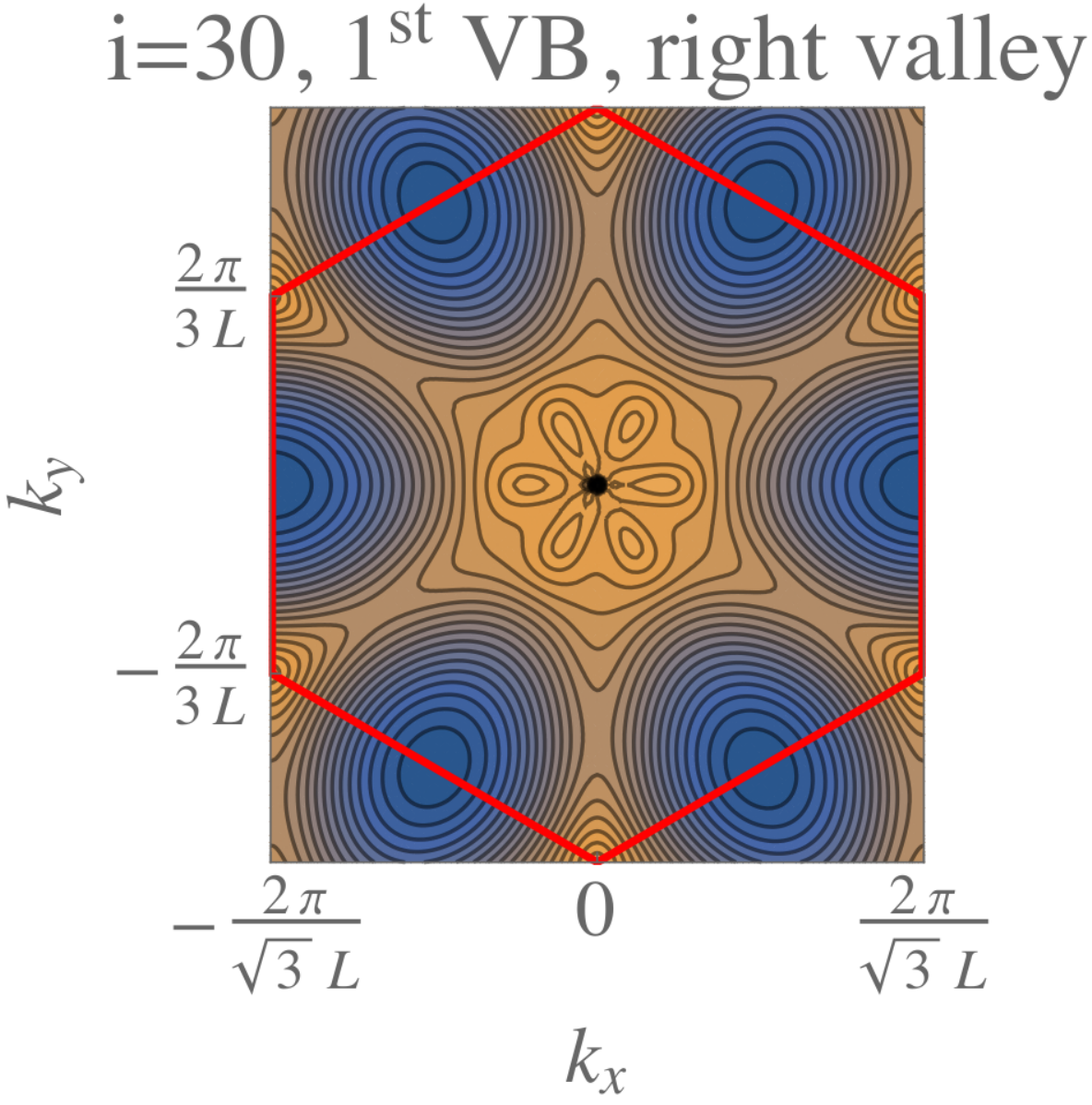}
\includegraphics[width=0.435\columnwidth]{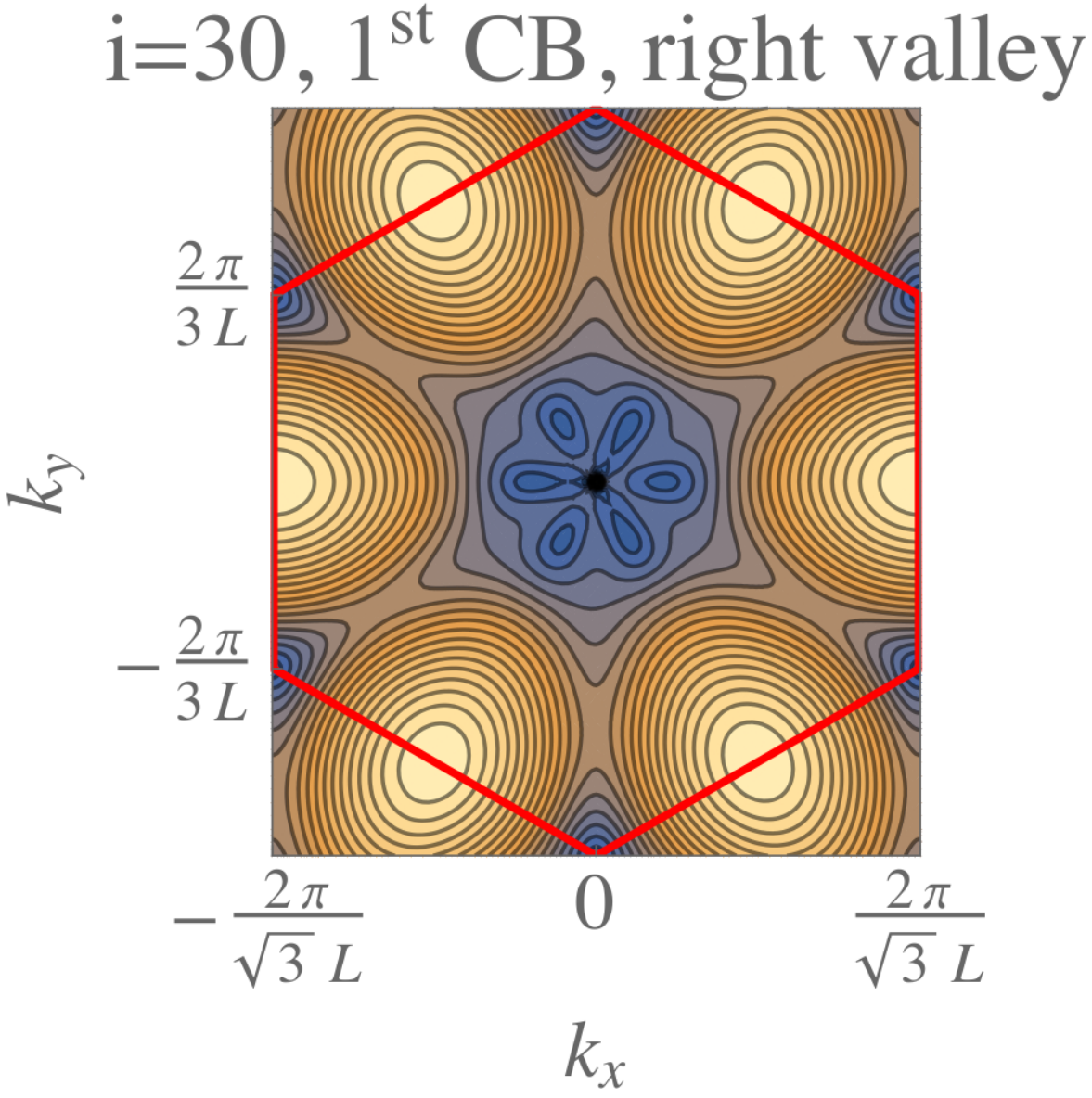}
\includegraphics[width=0.435\columnwidth]{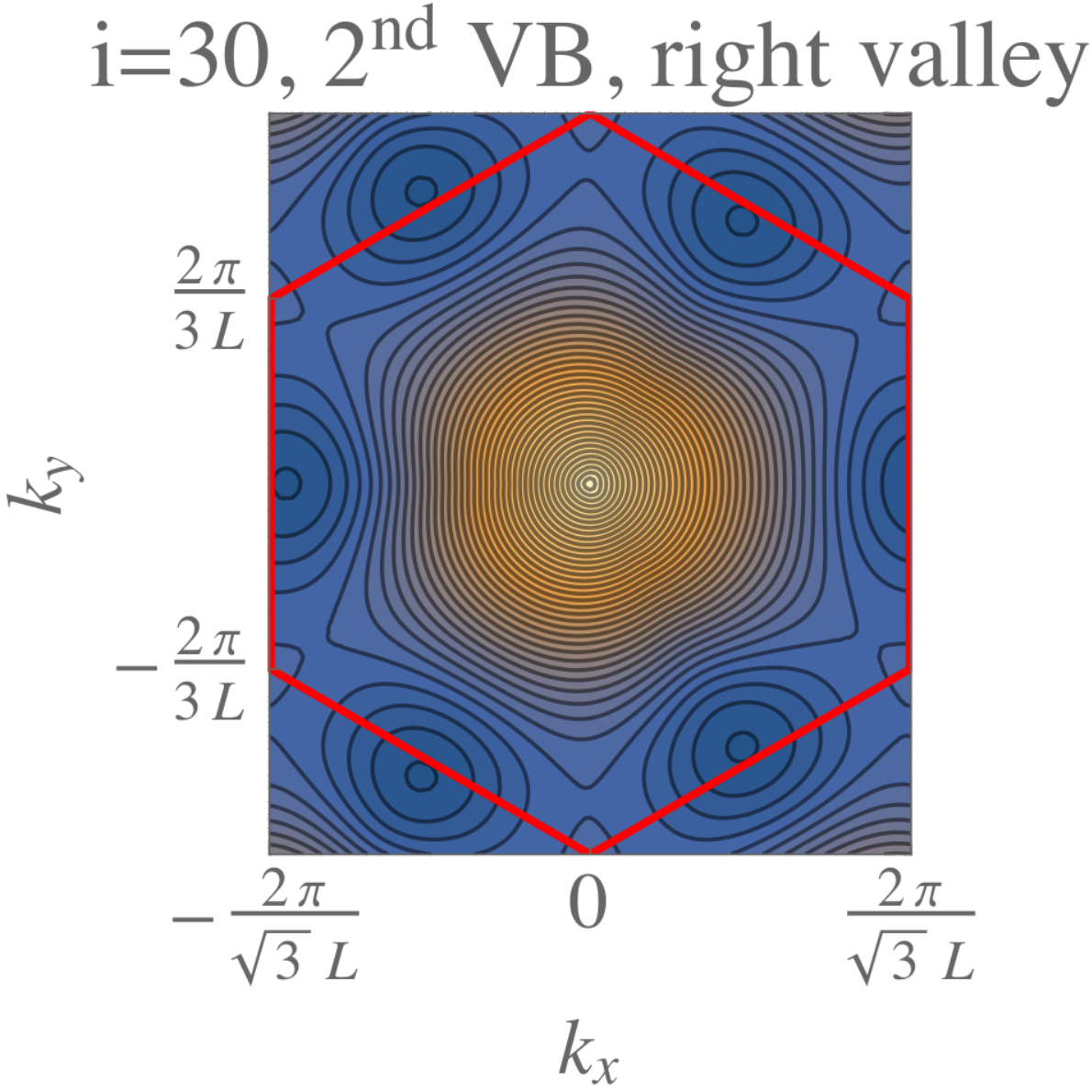}
\includegraphics[width=0.435\columnwidth]{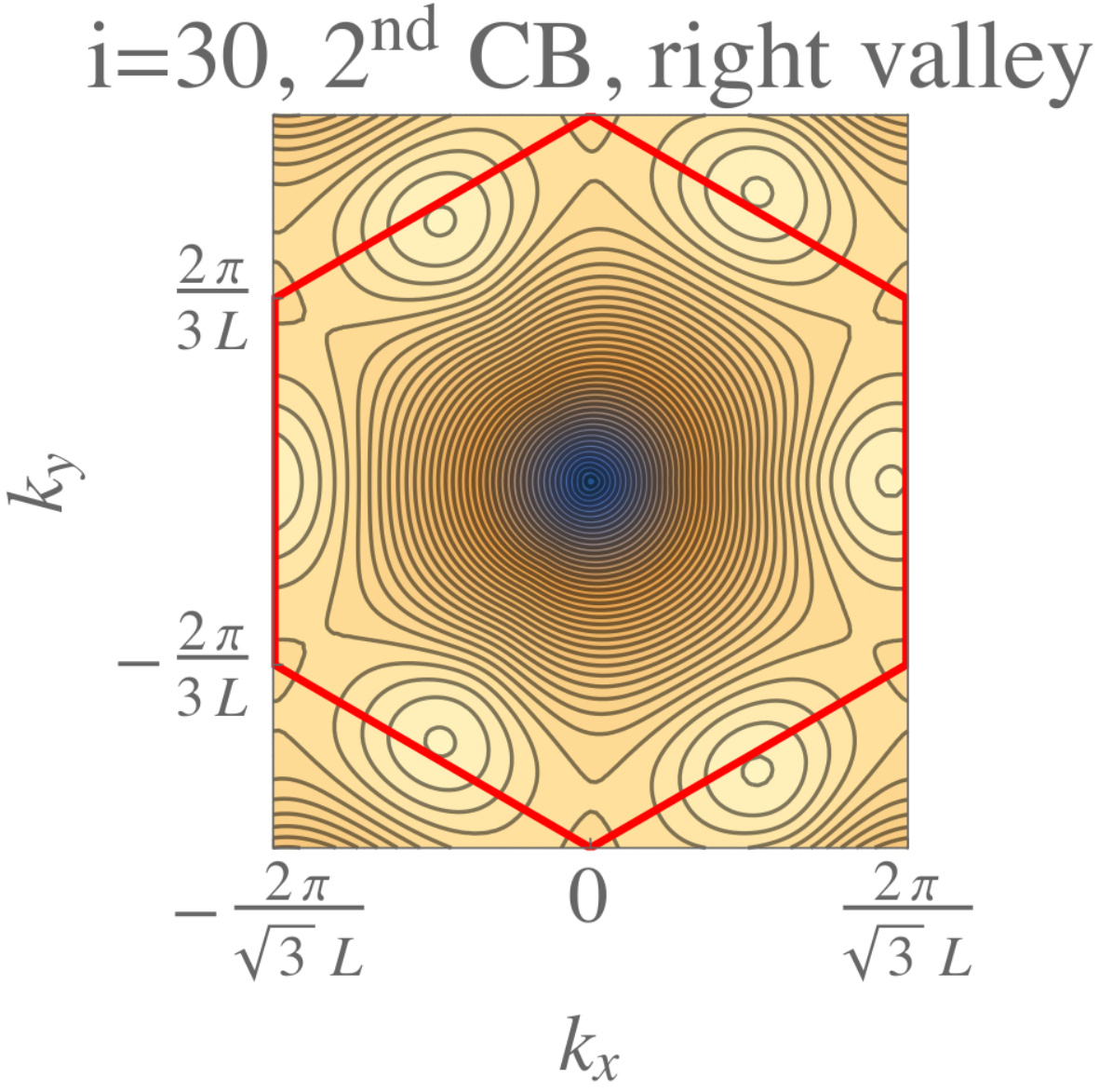}
\caption{{\it Upper panels:} Band structure of the first valence and conduction bands for the two valleys at $i=28$. The change in chirality is clear with respect to the first valence and conduction bands as well as with respect to the two valleys. {\it Lower panels:} The band structure of the first two valence and conduction bands for the twist angle at $i=30$. Notice that the emergent symmetry of $C_6$ is also present in the second valence and conduction band, respectively.}
\label{ph}
\end{figure*}
{\it Dual energy bands of TBG.} In order to gain more insight regarding the sign-change of $D_{xy}$, let us first discuss the spectrum close to the magic angle. The band structure of twisted bilayer graphene can be understood from the hybridization of the Dirac cones of the two carbon layers. Under the twist of the layers, their respective Brillouin zones (BZs) undergo a relative rotation, leading to a mismatch $\Delta \mathbf{K}$ of the Dirac nodes at the $K$ points. The low-energy bands of twisted bilayer are then folded onto two smaller Moir\'e-BZs which correspond to the two different valleys.

In the continuum model, the low-energy bands are obtained separately for each of the two independent Moir\'e-BZs. For one valley, we have the Hamiltonian
\begin{widetext}
\begin{eqnarray}
H_R=
v_F \left(
 \begin{array}{cccc}
0 &  -i\partial_x - \partial_y + i \Delta{\bf K}/2 & 
                           V_{AA'}(\mathbf{r}) & V_{AB'}(\mathbf{r}) \\
 -i\partial_x + \partial_y - i \Delta{\bf K}/2 & 0 & 
                           V_{BA'}(\mathbf{r}) & V_{AA'}(\mathbf{r}) \\
 V_{AA'}^\star(\mathbf{r}) & V_{BA'}^\star(\mathbf{r}) & 0 & 
                         -i\partial_x - \partial_y - i \Delta{\bf K}/2 \\
 V_{AB'}^\star(\mathbf{r}) & V_{AA'}^\star(\mathbf{r}) &  
                        -i\partial_x + \partial_y + i \Delta{\bf K}/2 & 0
 \end{array}\right)\;,
\label{H}
\end{eqnarray}
\end{widetext}
where $v_F$ is the Fermi velocity of graphene and $V_{AA'}({\bf r}), V_{AB'}({\bf r})$ and 
$V_{BA'}({\bf r})$ are the respective interlayer tunneling amplitudes between regions of stacking $AA, AB$, and $BA$. The Hamiltonian $H_L$ operating in the other Moir\'e-BZ is similar to $H_R$, with the only difference that the shift $\Delta \mathbf{K}$ is replaced by $-\Delta \mathbf{K}$ and the momentum $k_x$ is reversed after changing from one graphene valley to the other.

The low-energy bands $\varepsilon_L (\mathbf{k}), \varepsilon_R (\mathbf{k})$ obtained from the respective Hamiltonians $H_L, H_R$ do not have in general the same shape, but they are mapped onto each other by the transformation $k_x \rightarrow - k_x$. This corresponds to the fact that the lattice of the twisted bilayer remains invariant under a mirror-symmetry transformation about the $y$-axis in real space, accompanied by the exchange of the two carbon layers. In general, each of the bands $\varepsilon_L (\mathbf{k})$ and $\varepsilon_R (\mathbf{k})$ does not have separately such a mirror symmetry under the reversal $k_x \rightarrow - k_x$, as shown in the upper panels of Fig. \ref{ph}. There we also observe that the shapes of the valence and conduction bands are mapped onto each other under the dual transformation $k_x \rightarrow - k_x$.

Looking at the first valence and conduction bands, however, we see that there is a special twist angle at which the symmetry is enlarged, as the mirror symmetry $k_x \rightarrow - k_x$ becomes realized within each Moir\'e-BZ of the twisted bilayer, see lower panels of Fig. \ref{ph}. The value of such a critical twist angle is in the regime where the first magic angle is usually located, also discussed in the Supplemental Material of Ref. \cite{Gonzalez19}. For the parameters that we have used in our calculation with $t=2.78$eV and Fermi velocity $v_F = 4.2$ eV$\times a$, $a$ being the C-C distance, and interlayer coupling $w = 0.11$ eV as defined in Ref. \onlinecite{Bistritzer11}, the angle at which we find the enlarged symmetry corresponds to the index $i = 30$, in the sequence of commensurate lattices with $\theta_i = \arccos ((3i^2+3i+0.5)/(3i^2+3i+1))$. The approach to the critical angle is clearly noticed in the contour plots of the low-energy bands, as they are promoted to a $C_6$ symmetry which is in general absent within each Moir\'e-BZ.

\begin{figure}[t]
\includegraphics[width=0.99\columnwidth]{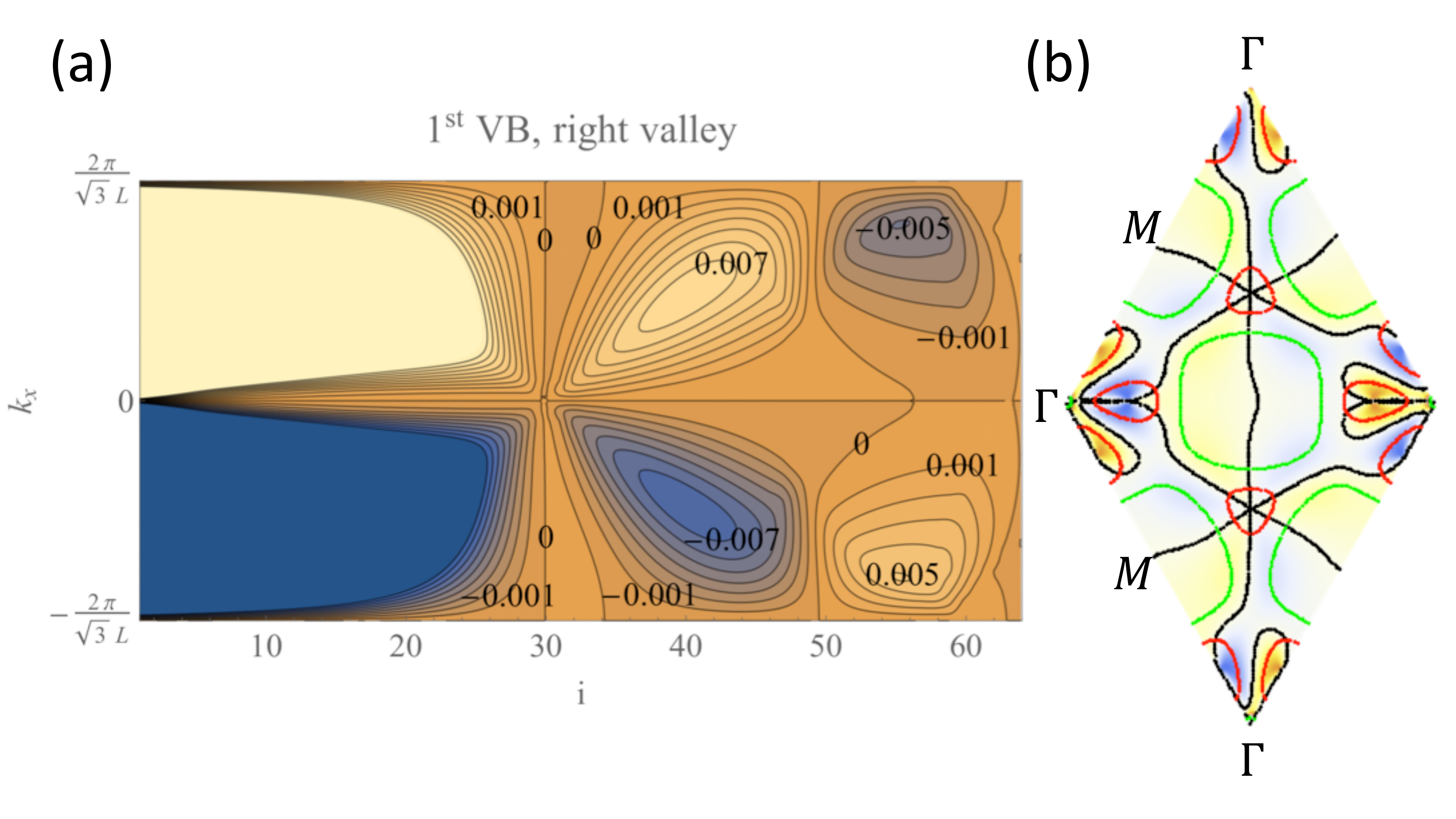}
\caption{(a) Plot of the deviation from mirror symmetry $\varepsilon (k_x,0) - \varepsilon (-k_x,0)$ of the highest valence band within a single valley of twisted bilayer graphene, as a function of the index $i$ in the sequence of commensurate superlattices. The vertical line at $i=30$ corresponds to the point of almost complete mirror symmetry realized at the first magic angle of twisted bilayer graphene, the rightmost arrow in Fig. \ref{Fig1} (b). Other approximate symmetry lines at $i=50$ and $i=63$ correspond to higher-oder magic angles, also marked by arrows in Fig. \ref{Fig1} (b). (b) Density plot of $\e_z\cdot(\j_{\k}^1\times\j_{\k}^2)$ of the first valence band at $i=30$ on the first Brillouin zone. The black line indicates the sign change of $\e_z\cdot(\j_{\k}^1\times\j_{\k}^2)$, the red and green lines denote the Fermi lines for two characteristic Fermi energies $E_F$.}
\label{index}
\end{figure}

We observe, therefore, that there is a critical angle at which an enhanced symmetry is realized in the twisted bilayer. This is indeed an emergent symmetry, since it is not implied by the symmetry transformations in the lattice of the bilayer. The location of the critical angle can be taken as a more precise identification of the magic angle, as this is usually referred to a situation where the first valence and conduction bands become flat, while in practice such a flatness is only approximate. In the case of the critical angle here described, the realization of the emergent mirror symmetry is, however, very precise, as seen in Fig. \ref{index} (a). Moreover, the location of the critical angle has a clear observational counterpart, since it corresponds to the point where the sense of chirality is lost in the twisted bilayer, in accordance with the enhanced symmetry of the low-energy bands. Also approximate symmetry lines at $i=50$ and $i=63$ correspond to sign changes in $D_{xy}$, as is indicated by the two left arrows of Fig. \ref{Fig1} (b).

{\it Chiral Drude weight.}
Even though the spectrum of two different enantiomers is identical, the change of chirality can be related to an emergent $C_6$-symmetry that is seen in the band  structure, i.e., usually, parity is broken in one valley and the bands only display a $C_3$-symmetry. Moreover, valence and conduction bands show dual $C_3$-symmetries which suggests a sign change at the neutrality point.  

However, chirality cannot be discussed using the band-structure alone, but only in connection with the eigenvalues as is clear from its definition, Eq. (\ref{HallDrude}). In the following, we will recall the introduction of the observable $D_{xy}$ based on linear response theory that characterizes the chirality of the electronic subsystem of twisted bilayer graphene.\cite{Stauber18,Stauber18b} As mentioned in the introduction, this chirality does not necessarily have to coincide with the chirality of the underlying atomic system. 

$D_{xy}$ is the low frequency limit of the current-current correlation function $\chi_{j_x^1,j_y^2}(\omega)=-\frac{i}{\hbar}\int_0^\infty dte^{i\omega t}\langle[j_x^1(t),j_y^2]\rangle$, i.e., $D_{xy}=$lim$_{\omega\to0}\chi_{j_x^1,j_y^2}(\omega)$. $D_{xy}$ can, therefore, be denoted as chiral or Hall Drude weight. This quantity needs to be contrasted with the total Drude weight defined by $D_{T}=$lim$_{\omega\to0}\chi_{j_i,j_i}(\omega)$ with the total current $j_i=j_i^1+j_i^2$ in direction $i=x,y$. 

Since the total Drude weight is related to the inverse mass tensor, it can be calculated from the knowledge of the band structure as $D_{T}=\frac{1}{A}\sum_{\k,n}\left(\frac{e}{\hbar}\frac{\partial\epsilon_{\k,n}}{\partial k_x}\right)^2\delta(\epsilon_{\k,n}-E_F)$. Interestingly, also the chiral Drude weight can be written in terms of a generalized density of states. This can be derived by relating the equilibrium response, $\chi_{j^aj^b}$, and the adiabatic Drude response, $D_{j^aj^b}$, via the contact term\cite{Stauber18b}
\begin{align}
\chi_{j^aj^b}=D_{j^aj^b}+\frac{1}{A}\sum_{\k,n}j_{\k,n}^aj_{\k,n}^b\delta(\epsilon_{\k,n}-E_F)\;.
\end{align}
Eq. (\ref{HallDrude}) now follows from gauge invariance,\cite{Stauber18b}
 i.e., from the absence of current-carrying states in equilibrium $\chi_{j_x^1,j_y^2}=0$,\cite{Bohm49} and rotational symmtery, i.e.,  $D_{j_x^1,j_y^2}=-D_{j_y^1,j_x^2}$. Its definition also implies that $D_{xy}$ is an odd function with respect to the twist angle $\theta$. For the particle-hole symmetric continuum model, one further has the exact symmetry $D_{xy}(E_F)=-D_{xy}(-E_F)$. The effect of temperature can be included by the usual replacement of the $\delta$-function by the derivative of the Fermi function. 

Let us recall that the Hall Drude weight gives rise to the longitudinal Hall effect,\cite{Stauber18} $\j=-aD_{xy}\B$, where $a\sim3.4$\AA$ $ is the distance between the two layers and $\B$ is the in-plane magnetic field causing the longitudinal current $\j$. This is not an equilibrium property, but the response to the adiabatic introduction of the field, thus accessible to slow driving transport experiments in clean samples.

The results for $D_{xy}$ as function of the Fermi energy are obtained from the continuum model of TBG\cite{Lopes07,Bistritzer11} and summarized in Fig. \ref{Fig1}. For large angles, there is a sign change at the neutrality point as the carrier type changes from holes to electrons. This is reminiscent of the ordinary Hall effect. But for the twist angle with $i=30$, i.e., $\theta_{30}=1.08^\circ$, we observe an abrupt change in chirality for Fermi energies up to $E_F\sim\pm75$meV corresponding to the second valence and conduction band, see rightmost arrow in Fig. \ref{Fig1} (b). This energy scale also coincides with the second van Hove singularity branch of the Density of States (DOS) at twist angle $\theta_{30}$ and a fractal structure develops in both the DOS and $D_{xy}$ for lower twist angles. For $D_{xy}$, there are further sign changes inside the dome which define higher-order magic angles, indicated by the two left arrows in Fig. \ref{Fig1} (b). In contrary to the first magic angle, these do not coincide with those obtained in Ref. \onlinecite{Bistritzer11}, but to an approximate $C_6$-symmtery as discussed in Fig. \ref{index} (a). 

Let us finally note that the chirality has been calculated for the continuum model with equal interlayer hopping in the $AA$- and $AB$-region at low temperature. At room temperature, the change in chirality around the magic angle is lost. Gradually reducing the hopping in the $AA$-region leads to a overall reduction of $D_{xy}$ until it vanishes for all twist angles at $t_{AA}=0$ for which the $C_6$-symmetry has been recovered.\cite{PhysRevLett.108.216802,Tarnopolsky19} 

\begin{figure}[t]
\includegraphics[width=0.99\columnwidth]{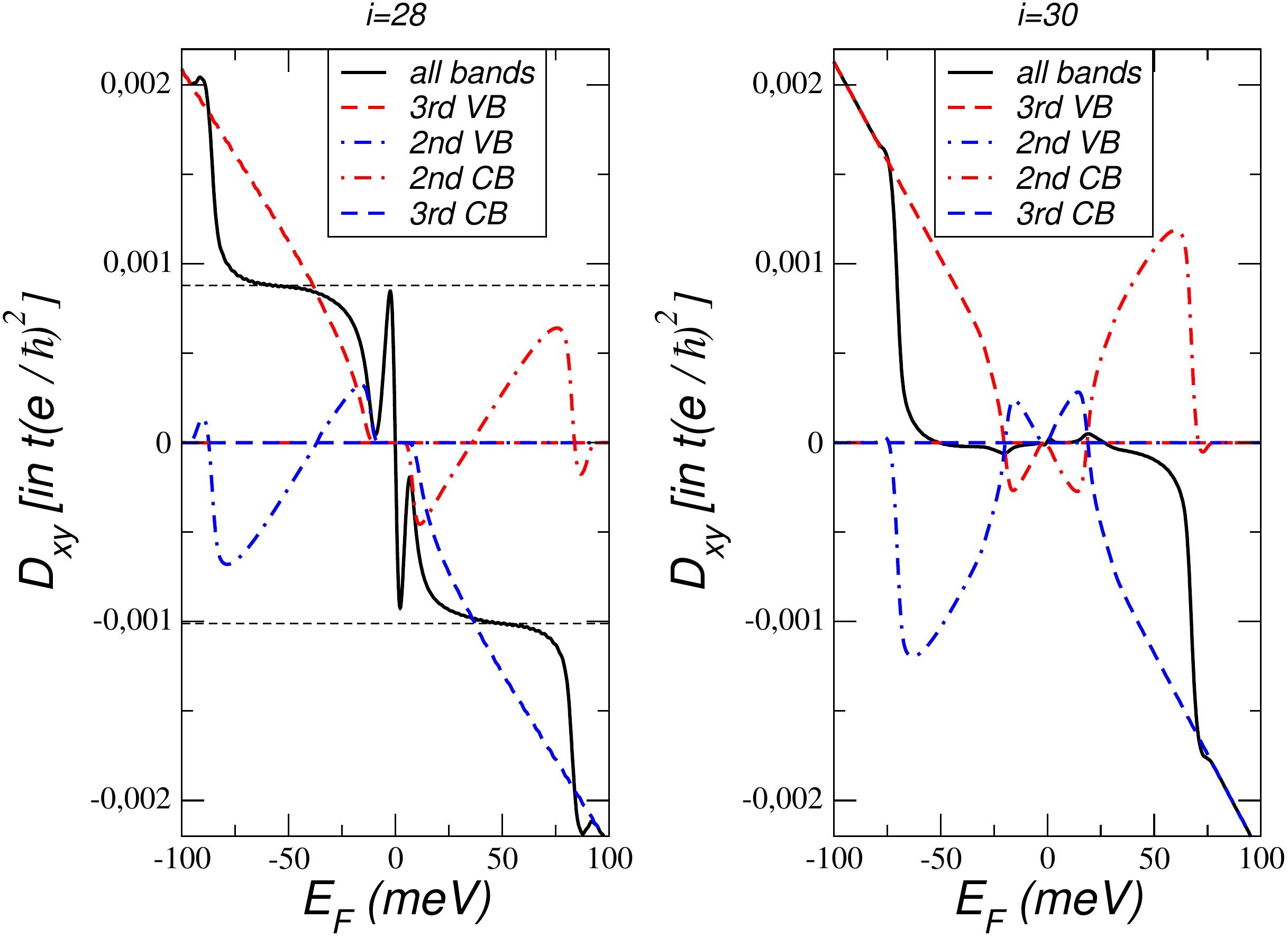}
\includegraphics[width=0.99\columnwidth]{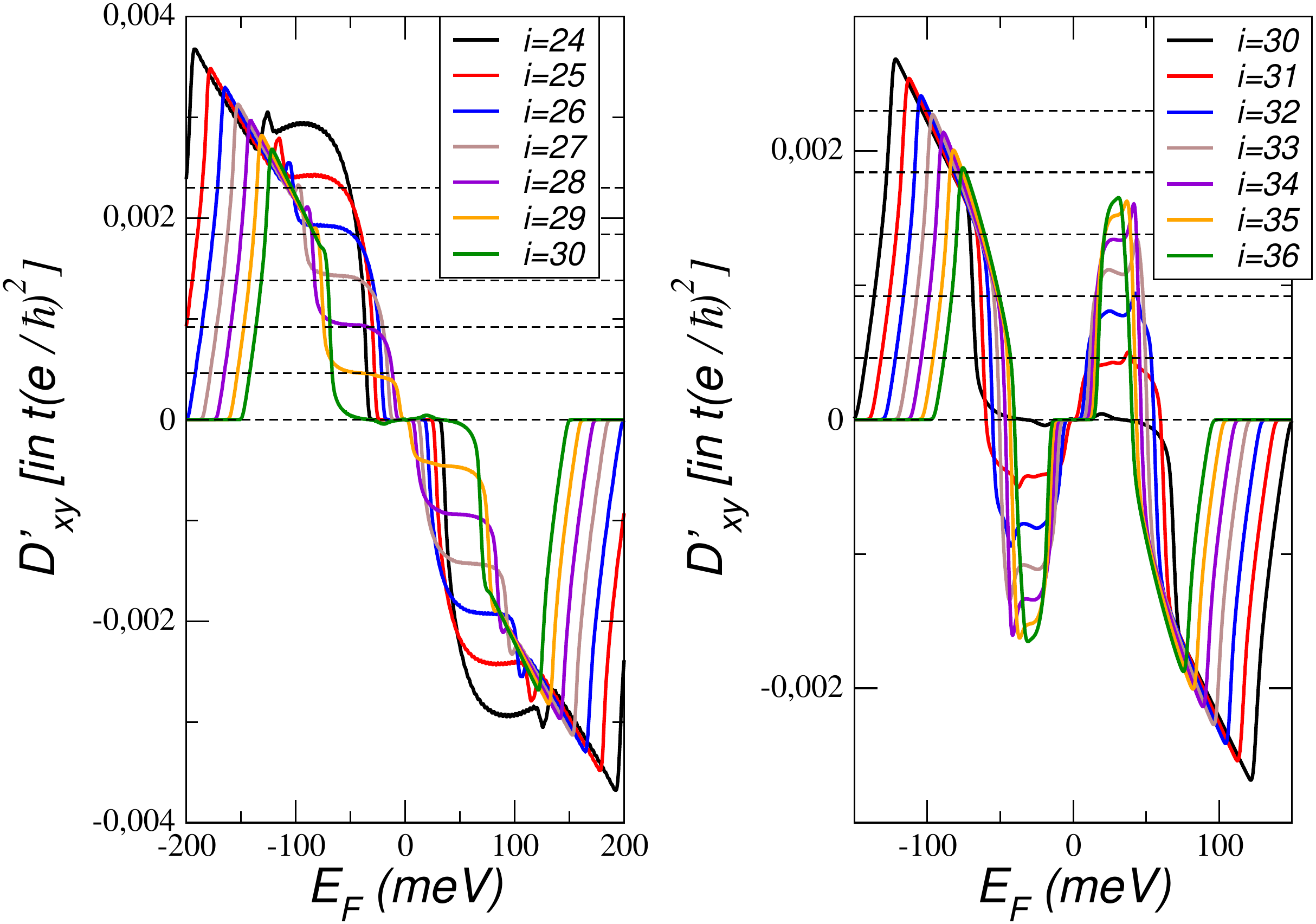}
\caption{{\it Upper panels:} The chirality $D_{xy}$ (black full line) as function of the Fermi energy for $i=28$ and $i=30$. Also shown the contributions of the 3rd VB (red dashed line), 2nd VB (blue dashed-dotted line), 2nd CB (red dashed-dotted line), 3rd CB (blue dashed line). {\it Lower panels:} $D_{xy}'$ for the electron-hole symmetric model that only contains the contribution from the 2nd and 3rd valence and conduction bands.
}
\label{DxyMagicAngle}
\end{figure}

{\it Quantized plateaus at remote bands.}
With an emergent $C_6$-symmetry, the chiral Drude weight vanishes for Fermi energies within the first valence band (VB) and conduction band (CB), respectively. The absence of $D_{xy}$ is due to a non-trivial cancellation of $\e_z\cdot(\j_{\k}^1\times\j_{\k}^2)$ when summing its contribution around the corresponding Fermi lines at fixed $E_F$. This can be seen in Fig. \ref{index} (b), where we show the density plot of this quantity for the first valence band at $i=30$ on the first Brillouin zone. The black line indicates the sign change of $\e_z\cdot(\j_{\k,n}^1\times\j_{\k,n}^2)$, the red and green lines denote the Fermi lines for two characteristic Fermi energies. 

Interestingly, there is also an almost perfect cancellation of $D_{xy}$ within the second VB and CB which comes from the interplay with the third VB and CB. Furthermore, for twist angles close to the magic angle both remote bands give rise to a plateau-like structure. This can be seen in the upper panels of Fig. \ref{DxyMagicAngle}. 

There is an approximate quantization of the plateau for commensurate twist angles with $i\lesssim30$. This is shown on the lower panels of Fig. \ref{DxyMagicAngle}, where the plateaus are almost  equally spaced for integer number $i$ with $\Delta D_{xy}=0.00046t\frac{e^2}{\hbar^2}$. For $i\gtrsim30$, this quantization becomes worse. For the sake of clarity, we plotted $D_{xy}'$ for the electron-hole symmetric model that only contains the contribution from 2nd and 3rd valence and conduction bands.

{\it Summary.} We have introduced a simple formula of a macroscopic quantity that defines the chirality of the electronic subsystem of TBG. This quantity changes sign due to an
emergent $C_6$ symmetry that occurs when the highest VBs as well as the
lowest CBs from the two valleys, both endowed with dual $C_3$ symmetries, merge and eventually cross at the magic angle. Interestingly, this happens for all wave numbers at approximately the same twist angle. Even more noteworthy, we find the absence of chirality on a much larger energy scale which naturally leads to a new definition of a magic angle, i.e., the angle for which the chirality of the system vanishes that should be detectable in in-plane magneto-resistance measurements. For commensurate angles greater than the magic angle, we find an approximate quantization of $D_{xy}$ for Fermi energies within the first remote band. Hopefully, our findings will give new insights in the underlying physics leading to the emergence of flat bands at magic angles. 

{\it Acknowlegdements.}
We thank L. Brey, F. Guinea, and J. Schliemann for interesting discussions. This work has been supported by Spain's MINECO under Grant No. FIS2017-82260-P, PGC2018-096955-B-C42, and CEX2018-000805-M as well as by the CSIC Research Platform on Quantum Technologies PTI-001. TS also acknowledges support from Spain's "Salvador de Madariaga"-Programme under Grant No. PRX19/00024 and from Germany's Deutsche Forschungsgemeinschaft (DFG) via SFB 1277.
\bibliography{Chirality.bib} 
\end{document}